\documentclass[12pt]{article}
\usepackage[psamsfonts]{amssymb}
\usepackage{cmmib57}
\usepackage{amsmath,amssymb,amscd}
\newcommand{\bPf}{\par\vspace*{-4pt}\indent{\sc Proof.}\enskip}
\newcommand{\ePf}{\medskip}
\def\QED{\hskip0.1em\hfill\null\ \null\nobreak\hfill\kern3pt\vbox{\hrule\hbox
   {\vrule\kern1pt\vbox{\kern1.7pt\hbox{$\scriptscriptstyle{QED}$}
    \kern0.2pt}\kern1pt\vrule}\hrule}}

\def\END{\hskip0.1em\hfill\null\ \null\nobreak\hfill\kern3pt\vbox{\hrule\hbox
   {\vrule\kern1pt\vbox{\kern1.7pt\hbox{$\,\,\,\vspace{5pt}$}
    \kern0.2pt}\kern1pt\vrule}\hrule}}
\newtheorem{theorem}{Theorem}
\newtheorem{lemma}{Lemma}
\newtheorem{corollary}{Corollary}
\newtheorem{proposition}{Proposition}
\newtheorem{remark}{Remark}
\newtheorem{definition}{Definition}
\newtheorem{example}{Example}

\newcommand{\bCd}{\bEq\begin{CD}}
\newcommand{\eCd}{\end{CD}\eEq}
\newcommand{\bcd}{\beq\begin{CD}}
\newcommand{\ecd}{\end{CD}\eeq}
\newcommand{\ben}{\begin{enumerate}}
\newcommand{\een}{\end{enumerate}}
\newcommand{\bEq}{\begin{eqnarray}}
\newcommand{\eEq}{\end{eqnarray}}
\newcommand{\beq}{\begin{eqnarray*}}
\newcommand{\eeq}{\end{eqnarray*}}
\newcommand{\bDf}{\begin{definition}\em}
\newcommand{\eDf}{\end{definition}}
\newcommand{\bLm}{\begin{lemma}}
\newcommand{\eLm}{\end{lemma}}
\newcommand{\bPr}{\begin{proposition}}
\newcommand{\ePr}{\end{proposition}}
\newcommand{\bTh}{\begin{theorem}}
\newcommand{\eTh}{\end{theorem}}
\newcommand{\bCr}{\begin{corollary}}
\newcommand{\eCr}{\end{corollary}}
\newcommand{\bRm}{\begin{remark}\em}
\newcommand{\eRm}{\end{remark}}
\newcommand{\bEx}{\begin{example}\em}
\newcommand{\eEx}{\end{example}}

\newcommand{\Z}{\mathbb{Z}}
\newcommand{\ie}{{\em i.e$.$} }
\newcommand{\eg}{{\em e.g$.$} }
\newcommand{\R}{I\!\!R}

\newcommand{\der}{\partial}

\newcommand{\cE}{\mathcal{E}}

\newcommand{\cL}{\mathcal{L}}

\newcommand{\cP}{\mathcal{P}}

\newcommand{\bG}{\boldsymbol{G}}

\newcommand{\bK}{\boldsymbol{K}}

\newcommand{\bP}{\boldsymbol{P}}

\newcommand{\bR}{\boldsymbol{R}}

\newcommand{\bU}{\boldsymbol{U}}

\newcommand{\bX}{\boldsymbol{X}}
\newcommand{\bY}{\boldsymbol{Y}}

\newcommand{\sub}{\subset}

\newcommand{\ten}{\!\otimes\!}

\newcommand{\del}{\delta}
\newcommand{\eps}{\epsilon}

\newcommand{\lam}{\lambda}
\newcommand{\sig}{\sigma}

\newcommand{\For}{{\Lambda}}

\newcommand{\Var}{{\mathcal{V}}}
\newcommand{\Thd}{{\theta}}
\title{\large{\bf A cohomological obstruction in higher dimensional Chern--Simons gauge theories}}
\author{{\normalsize Marcella Palese and Ekkehart Winterroth\footnote{Corresponding author.}}
\\
{\footnotesize Department of Mathematics, University of Torino, Italy}
\\ 
{\footnotesize via C. Alberto 10, 10123 Torino, Italy}
\\
{\footnotesize e--mail: {\sc marcella.palese@unito.it, ekkehart.winterroth@unito.it}}
}
\date{}
\begin{document}
\maketitle

\begin{abstract}
We study a set of cohomology classes which emerge in the cohomological formulations of the calculus of variations as obstructions to the existence of (global) solutions of the Euler--Lagrange equations of Chern--Simons gauge theories in higher dimensions $2p+1 > 3$. 
\end{abstract}



\section{Introduction}

In the differential (or geometric) formulation of the calculus of variations, `taking the variation' of a Lagrangian corresponds to taking the differential of a cochain, the Lagrangian, in some complex, a particular kind of quotient of the de Rham complex of the total space of a fibre bundle. The first fundamental result in this context was the solution of the so called `inverse problem' of the calculus of variations, \ie the questions as to when equations are Euler-Langrange equations, \ie derive from the variation of a Lagrangian.

The solution splits into two parts. A certain type of cochain, a so called {\em dynamical} or {\em source} form, is {\em locally} variational if it is a cocycle, \ie it is closed in the complex. Then, there exists at least a $1$-cocycle of local Lagrangians.
This cochain is {\em globally} variational, \ie the $1$-cocycle can be replaced by a single global Lagrangian, if its cohomology class vanishes.

At this point it is natural to enlarge the concept of variational problem and to consider also `local variational problems', \ie the above $1$-cocycles of local Langrangians. This may be advantageous even when a globlal Lagrangian exists, since it may be hard to find, to write down or generally more cumbersome to deal with: Chern--Simons gauge theories are a well fitting example.

For local variational problems, global conservation laws can no longer be derived from the symmetries of the 
Lagrangians, instead one is lead to study symmetries of the corresponding dynamical form (which we assume to describe physically meaning {\em global } equations) within a combination of the first and second Noether theorem (this is sometime also called a Noether--Bessel-Hagen approach). 

While the conservation laws so derived remain global, the conserved quantities may not. There appears a topological obstruction, a cohomology class in the complex, which may not vanish, even if a global Lagrangian exists; see \cite{FePaWi10,FrPaWi13,PaRoWiMu16,PaWi16}.

We proved that this obstruction constitutes a link between the existence of global conserved quantities and the existence of global solutions: the pullback of the cohomology class to the base -and under certain conditions the class itself- is also an obstruction to the existence of global solutions \cite{PaWi16}.
We can study this cohomology class in various important special cases and (classical) Chern Simons gauge theories are a natural choice.
The $3$-dimensional case has been dealt with previously \cite{PaRoWiMu16}, \cite{PaWi16}. There, the obstruction proved to be sharp. 

In this paper we extend the results of \cite{PaWi16} to the $2p+1$-dimensional case for the $(p+1)$th-Chern polynomial and the unitary group $U(n)$, $n \geq p+1$, with $p$ arbitrary. The result is that the obstruction vanishes if and only if the $p$-th Chern class vanishes. However, it is not clear if the obstruction will still be sharp. It seems more likely that it is in this case a kind of primary obstruction, analogous to the characteristic classes in obstruction theory.

The way the obstruction arises shows the power of the cohomological formulations of the calculus of variations: the obstruction we characterize not only allows to identify de Rham cohomology classes as obstruction to the existence of solutions, but it can do so by means intrinsic to the calculus of variations. 
This is particularly relevant in theories involving the coupling of the Chern--Simons Lagrangian with another Lagrangian, where it may be used to derive non existence theorems in theoretical physics applications \cite{WYMCS}.
Furthermore, since our obstruction arises as an obstructions to the existence of global conserved quantities, we identify the the $p$th real Chern class of $\bP$ as an obstruction to the globality of conserved quantities.

There is quite some interest in higher Chern--Simons theories in theoretical physics (a sample selection spanning over three decades:  the seminal and well known \cite{Ch89}, \cite{Ch90}, 
the excellent review \cite{Zan12}; see also the recent \cite{BaVeCa21}) and it should be, thus, interesting to investigate the consequences of our findings.

Since, we touch upon various and quite disparate topics, we aim to give a reasonably self contained exposition.
We will first sketch how to construct a suitable complex for the differential approach to the calculus of variations (for a more thorough treatment see \eg \cite{Kru90}  for the construction on  finite order jet bundles). Note that, instead of using the contact splitting of $\pi^{r \,*}_{r-1}(T^*J_{r-1}\bY)$, starting from an analogous splitting one can work on jets of infinite order, see \eg \cite{And92,AnDu80,Tak79,Tul77,Tu88,Vin77}; our results remain valid also in the infinite order case.
Afterwards the concept of local variational problem is explained and the corresponding `Noether theorems' and the obstruction to the
existence of global conserved currents are introduced. 
Next the link with the existence of global solutions is established. We then  deal first with the formulation of Chern--Simons theories on the bundle of connections before investigating the conditions of existence of global solutions and the role of (variational) cohomology  therein in the specific case of the $p+1$ Chern polynomial and the unitary group $U(n)$,  $n \geq p+1$. 

\section{Local variational problems, symmetries and conservation laws}

In this section we sketch some of the fundamental constructions and classical results of the differential formulation of the calculus of variations; see \eg \cite{Kru90,Saund}. 
We assume the $r$-th order prolongation of a fibered manifold $\pi: \bY \to \bX$, with $\dim \bX = n$ and $\dim \bY = n+m$, to be the configuration space; \ie  fields are  (local) sections  of $\pi^{r}: J_r \bY \to \bX$.
The affine bundle structure of  
$\pi^{r+1}_{r}: J_{r+1} \bY \to J_{r} \bY$, induces a natural splitting
$J_r\bY\times_{J_{r-1}\bY}T^*J_{r-1}\bY =
J_r\bY\times_{J_{r-1}\bY}(T^*\bX \oplus V^*J_{r-1}\bY)$, and therefore corresponding natural  {\em splittings} in  horizontal and vertical parts of vector fields, forms and of the exterior differential on $J_r\bY$.
Starting from this splitting one can define sheaves of contact forms $\Thd^{*}_{r}$ defined by the kernel of the horizontalization. The local generators of the contact ideal are well known pfaffian $1$-forms defining higher order partial derivatives, and their differentials.
The sheaves $\Thd^{*}_{r}$ form an exact subsequence of the de Rham sequence on $J_r\bY$ and one can define the quotient sequence
$\{ 0\to \R_{\bY} \to \For^{*}_r / \Thd^{*}_r \,, \cE_{*} \}$, 
called the $r$--th order {\em variational sequence} on $\bY\to\bX$, which is an acyclic sheaf resolution of the constant sheaf $\R_{\bY}$; see \cite{Kru90}, where it is also proved that the cohomology of the complex of global sections $H^{*}_{VS}(\bY)$ is naturally isomorphic to the de Rham cohomology $H^{*}_{dR}(\bY)$. Then it is also isomorphic to the \v Cech cohomology of $\bY$. 
Thus if the cohomology of $\bY$ is trivial, of course each local inverse problem is also global.

The quotient sheaves in the variational sequence can be represented as sheaves $\Var^{k}_{r}$ of $k$-forms on jet spaces of higher order, see \eg \cite{Kru90,PaRoWiMu16}.
Lagrangians are sheaf sections  $\lam\in \Var^{n}_{r}$, while $\cE_n$ is called the Euler--Lagrange morphism; the latter is thus characterized as a quotient morphism of the exterior differential morphism of the de Rham complex. Therefore, the de Rham cohomology $H^{*}_{dR}\bY$ of $Y$ appears naturally as a set of obstructions to globality in the calculus of variation.

The Euler--Lagrange equations  are therefore given by  $\cE_{n}(\lam)\circ j_{2r+1}\sig =0$ for (local) sections $\sig: \bX \to \bY$.
Sections $\eta\in\Var^{n+1}_{r}$ are called {\em source forms} or also {\em dynamical forms}, 
while $\cE_{n+1}$ is called the Helmholtz morphism. 

In the case of a nontrivial cohomology of $\bY$, given a closed section of a quotient sheaf of the variational sequence, one look at the problem as to when this section is also globally exact. 
 
To answer to this question, let $\bK^{p}_{r}\doteq \textstyle{Ker}\,\cE_{p}$; we have a natural short exact sequence of sheaves
which  gives rise in a standard way to a long exact
sequence in \v Cech cohomology,
where the {\em connecting homomorphism}, explicitly given  by  $\delta_{p} = i^{-1}\circ\mathfrak{d}\circ\mathcal{E}_{p}^{-1}$, is the mapping of cohomologies in the corresponding diagram of cochain complexes ($\mathfrak{d}$ is the {\em coboundary operator} in these complexes). 

Every global section $\eta\in\cE_{p}(\Var^{p}_{r})$, \ie locally variational, defines a cohomology class 
$\del_{p} (\eta) \in
H^{1}(\bY, \bK^{p}_{r}) $ $\simeq$ $ H^{p+1}_{VS}(\bY)
$ $\simeq$ $ H^{p+1}_{dR}(\bY)$ (here $\simeq$ denotes the natural isomorphism between cohomologies).

Every non vanishing cohomology class in  $H^{p}_{dR}(\bY)$ gives rise to local variational problems.
It is clear that $\eta$ is globally variational if and only if $\del_{p} (\eta) = 0$ (solution to the so called {\em global inverse problem}). 

Let us consider the case when, instead, 
$ [\eta]\simeq 
\del_{p} (\eta)  \neq 0$; then $\eta =  \cE_n(\lam)$ can be solved only locally, \ie for any countable good covering $\{\bU_{i}\}_{i\in \Z}$ in $\bY$ there exist local Lagrangians $\lam_{i}$ over each subset $\bU_{i}\sub\bY$ with $\eta_{i}=\cE_{n}(\lam_{i})$.
The local Lagrangians satisfy  $\cE_{n}((\lam_{i}-\lam_{j})|_{U_{i}\cap U_{j}}) = 0$ and conversely  any system of local sections of with this property gives rise to a Euler--Lagrange morphism $\eta\in(\cE_{n}(\Var^{n}_{r}))_{\bY}$ with cohomology class  $[\eta] \in
H^{n+1}(\bY, \R)$. 

Such a system of local sections  $\lam_{i}$ of  $(\Var^{n}_{r})_{U_{i}}$ for an arbitrary covering  $\{\bU_{i}\}_{i\in \Z}$ in $\bY$ is what we call a  {\em local variational problem}. Two local variational problems are {\em equivalent } if and only if they give rise to the same Euler--Lagrange morphism.

Note, in particular, that {\em we do not exclude the possibility} $[\eta] = 0$; \ie we allow also problems which admit a global Lagrangian as local variational problems. A system of local Lagrangians may be the natural way in which a variational problem presents itself and a global Lagrangian, if it exists, may be hard to find and difficult to deal with - Chern--Simons theories are an example of this situation. 

To obtain conservations laws, it is now natural to consider the symmetries of $\eta$, which is assumed to be a global object, instead of those of the Lagrangian, which could be or present itself as a local object. Cohomology enters in globality problems concerned with conserved quantities, we shall examine this aspect within Noether formalism \cite{Noe18}. 

Factorizing modulo contact structures 
 is also the basic idea underlying the definition of  a 
{\em variational Lie derivative} operator $\cL_{j_{r}\Xi}$ and of a {\em variation  formula} defined on the sheaves of the variational sequence. This enables to define symmetries of classes in the variational sequence and corresponding (eventually higher) conservation theorems; see \cite{AcPa21,CaPaWi16,PaRoWiMu16}.
The corresponding `Cartan formula' for the variational Lie derivative of closed classes of forms selects a quite important class defined by both the vertical part of the symmetry and the Euler--Lagrange class, as a consequence of  $\del_p (\cL_{\Xi}\eta_{\lam}) = 0$, \ie of the fact that the variational Lie derivative `trivializes' cohomology classes, see \eg \cite{FePaWi10,FrPaWi13}.

Let then now $\eta_{\lam}$ denotes a global Euler--Lagrange morphism for a local variational problem $\lam_i$; 
if $\eps_i $ $=$ $j_{r}\Xi_{V}\rfloor p_{d_{V}\lam_{i}}+\xi\rfloor\lam_{i}$ denotes a local canonical Noether current, for a projectable vector field $\Xi$, Noether's First Theorem reads {\em locally} 
$\cL_{\Xi} \lam_{i}= \Xi_V\rfloor\eta_\lam +d_H \eps_i$. 
By a (global) symmetry $\Xi$, we have $\cL_{j_{r}\Xi} \eta=0 $, and we get {\em along critical sections}  (\ie solutions of the Euler--Lagrange equations) the  global conservation law
\beq
 0 = d_{H}\eps_{i}- \cL_{\Xi} \lam_{i},
\eeq
since $d_{H}\eps_{i}- \cL_{\Xi} \lam_{i}$ is just a local expression for $-\Xi_{V} \rfloor \eta$. 

However, this leads now to the rather curious situation of a global conservation law which may not admit global conserved quantities. 

Indeed, since we assume $\eta$ is locally variational, \ie $\cE_{n+1}$-closed, and  since $\cL_{j_{r}\Xi} \eta=0$  then 
\beq  
0=\cE_{n}(\Xi_{V}\rfloor \eta) \,;
\eeq 
therefore {\em locally} 
$\Xi_V\rfloor\eta_\lam =d_H\nu_i$. Thus, $\Xi_{V} \rfloor \eta_{\lam}$ defines a cohomology class and   
 to admit global conserved quantities this cohomology class has to vanish. 

We stress that, for a global symmetry, $\mathfrak{d}(\Xi_V\rfloor \eta_{\lam}) = 0$, \ie this contraction is globally defined, but in general we could have the obstruction 
$\del_{n-1}(\Xi_V\rfloor \eta_{\lam})\neq 0$, 
so that the current  $\nu_i$ would be a necessarily {\em local} object (conserved along the solutions of Euler--Lagrange equations).

On the other hand, {\em and  independently} (see \cite{Noe18,BeHa21}), for a generalized symmetry (a symmetry of $\eta$ but not of $\lam$), we get {\em locally}
\beq
\cL_{\Xi} \lam_{i}=d_H \beta_{i}\,,
\eeq
thus  we can write
$\Xi_{V} \rfloor \eta_{\lam}  + d_{H}( \eps_i  -  \beta_{i} )$  $=$ $0$. 
\bDf
We call the (local) current $\eps_{i} - \beta_{i}$ a {\em Noether--Bessel-Hagen current}. 
\eDf 
Note that a Noether--Bessel-Hagen current is the difference between the canonical Noether current (from the invariance of a Lagrangian) and 
the Bessel-Hagen current (from the invariance of equations). In general, the obstruction to the globality of the single type of current is given by different and, in principle, independent cohomolgy classes \cite{FePaWi10,FrPaWi13}.

Interesting application of Noether--Bessel-Hagen currents can be recognized \eg in the study of dynamical systems, such as mechanical systems where external forces are present, or cosmological system derived from scalar-tensor gravity with unknown scalar-field potential \cite{UrBaCa20}.

Under certain conditions a Noether--Bessel-Hagen current, associated with a generalized symmetry, turns out to be variationally equivalent to a Noether current (exact on-shell and generating a canonical conserved quantity) for a certain correspondingly invariant Lagrangian \cite{CaPaWi16}.

We stress that, by the Noether--Bessel-Hagen theorem, $\del_{n-1}(\Xi_V\rfloor \eta_{\lam})\neq 0$ is of course also an obstruction for the globality of this latter current (which is variationally equivalent to the local current $\nu_i$). In other words, a local Noether--Bessel-Hagen current can be globalized if and only if
$0 =  \del_{n-1}(\Xi_V\rfloor \eta_{\lam}) \simeq [\Xi_{V} \rfloor \eta_\lam]  \in H^{n}_{dR}(\bY)$ \cite{FePaWi10,FrPaWi13}.

\section{A cohomological obstruction to the existence of global critical sections}

The cohomology class $[\Xi_{V} \rfloor \eta]$ establishes a curious link between the existence of global conserved quantities and the existence of global critical sections. For this link to be meaningful, we have to require $\bX$ to be closed, \ie compact without boundary. To apply these results to non compact manifolds the existence of a suitable compactification is needed.

\bTh \label{Th1}
Let $\eta$  be the dynamical form of a local variational problem and let $H^{n}_{dR}(\bY) \sim  \pi^{*} (H^{n}_{dR}(\bX))$. If the variational problem admits (global) critical sections then all conservation laws derived from symmetries of the global field equations admit global conserved quantities.
\eTh
This theorem is a consequence of the fact the the class $[\Xi_{V} \rfloor \eta]$ vanishes along critical sections, see \cite{PaWi16}, and, thus, does not depend on the differential formulation of the calculus of variation one works with.

The condition  $H^{n}_{dR}(\bY) \sim  \pi^{*} (H^{n}_{dR}(\bX))$ is satisfied, for example, by all theories on vector or affines bundle. In particular, it is satisfied by all theories which can be formulated on the bundle of connections (see below), like Yang--Mills type or Chern--Simons type theories.

Obviously, if critical sections annihilate $[\Xi_{V} \rfloor \eta]$, then the non vanishing of this class (or one of its pullbacks) is an obstruction to the existence of global critical sections. 
More precisely, if the class $j \sig^{*}([ \Xi_{V} \rfloor \eta])$ does not vanish, neither $\sigma$ nor any section homotopical to it, \ie any deformation or variation of $\sigma$, can be critical. Thus, $j \sig^{*}([ \Xi_{V} \rfloor \eta])$ is an obstruction to the existence of global solutions for the (local) variational problem defined by $\eta$ in the homotopy class of $\sigma$. 

If we require $\pi$ and, thus, $j \sig$ again to induce isomorphisms between  
$H^{n}_{dR}(\bX)$ and  $H^{n}_{dR}(J\bY) \equiv H^{n}_{dR}(\bY) $,
the class $j \sig^{*}([ \Xi_{V} \rfloor \eta])$ vanishes if and only if $[\Xi_{V} \rfloor \eta]$ vanishes. 
\\Therefore, we have \cite{PaWi16}
\bCr \label{Cr1}
Let $\sigma$ be a section of $\bY$ over $\bX$. 
If $0 \neq j \sig^{*}([ \Xi_{V} \rfloor \eta]) \in H^{n}(\bX)$, then there is no (global) critical section in the homotopy class of $\sigma$. 

If $H^{n}_{dR}(\bY) \sim  \pi^{*} (H^{n}_{dR}(\bX))$, as before, there are no (global) critical sections if $[ \Xi_{V} \rfloor \eta] \neq 0$.
\eCr

On the one hand, Euler--Lagrange equations are essentially a set of partial differential equations and for partial differential equations even the existence of local solutions is a non trivial problem, involving deep analysis. From that point of view, one would think that topological obstructions are too coarse to play a 
particularly important role. On the other hand, we have seen already two classical results (on the global inverse problem and on the existence of global conserved currents), where a certain cohomology class is the only obstruction to globality.

Of course, it is actually a slight abuse of language to talk of {\em one} cohomology class as obstruction, since it depends on the symmetry as well.

Unfortunately, the class $[\Xi_{V} \rfloor \eta]$ seems almost impossible to study in general. Partly this is due to the difficulties with giving an explicit, but general isomorphism between variational cohomology  and de Rham cohomology; an important improvement in this respect would be to be able to keep track of the contact structure in the Leray--Serre spectral sequence. Even more important is that the definition of $[\Xi_{V} \rfloor \eta]$ involves a contraction and the contraction of a form with a vector field behaves very badly with cohomology.

Thus, one is naturally led to study $[\Xi_{V} \rfloor \eta]$ in interesting special cases and Chern--Simons theories are for their `topological' character the first natural choice. The three dimensional case was dealt with in in detail in \cite{PaWi16}. In the next section we investigate Chern--Simons theories in arbitrary (odd) dimensions and present some preliminary relevant results.

\section{Chern--Simons theories in dimension $2p+1$}

Classical Chern--Simons theory is a classical field theory for principal connections on an arbitrary principal bundle $P$ over an odd dimensional manifold $\bX$. In what follows, the dimension of $\bX$ will be  $2p+1$. We will be mainly concerned with $U(n)$, $n \geq p+1$, as structure group and the $k$th-Chern polynomial, though at the end we will shortly discuss some other cases. Most of the expository material will nevertheless be valid in general. Anyhow, it will be stated explicitly when we deal exclusively with our main case of interest. $\bX$ will still be required to be closed.

\subsection{The bundle of connections}

To study Chern--Simons theories with the variational sequence or, more generally, any differential formulation of the calculus of variations, it is necessary to describe connections as sections of a bundle. To do so we need the bundle of connections: for an arbitrary principle bundle $\bP$ it is defined as 
\beq
\mathcal{C}_{\bP} := J^{1}\bP/\bG \mapsto \bP/\bG \sim \bX\,,
\eeq
 this bundle, $\pi_{\mathcal{C}_{\bP}}: \mathcal{C}_{\bP} \mapsto \bX$, is an affine bundle modeled on the vector bundle 
\beq
T^* \bX \ten  V\bP/\bG \mapsto \bX,
\eeq
\ie the bundle of $V\bP/\bG$ valued $1$-forms on $\bX$. Locally $\mathcal{C}_{\bP}$ can be trivialized as $\mathcal{C}_{\bP}|_{U} \sim U \times \bR^{n} \ten \mathfrak{g}$ with $\mathfrak{g}$ the Lie algebra of $\bG$. If $(x^{i})_{1 \leq i \leq n}$ are coordinates on $\bX$ and $\mathfrak{g}_{p}$ is a base of  $\mathfrak{g}$, we have coordinates $(x^{i}, A^{p}_{i})$ ($A^{p}_{i}$ is the coefficient of the component $dx^{i} \ten \mathfrak{g}_{p}$) on $\mathcal{C}_{\bP}|_{U}$.

 If $h_{UV}: U \cap V \mapsto \bG$ is the transition function of the change of trivialisation from $\bP|_V$ to $\bP|_U$, 
\beq
 (x, \omega_{U}(x)) = (x, ad(h_{UV}(x)^{-1})(\omega_{V}(x)) + dh_{UV}(x)h_{UV}^{-1}(x))
\eeq
is the change of trivialisation from $\mathcal{C}_{\bP}|_{V}$ to $\mathcal{C}_{\bP}|_{U}$.

Thus, the principal connections on $P$ are in one to one correspondence with the (global) sections $\sigma$ of  the bundle $\pi_{\mathcal{C}_{\bP}}: \mathcal{C}_{\bP} \mapsto \bX$ in the following sense: 
every section defines the set of local connection forms on $\bX$ corresponding to a connection on $\bP$.
Furthermore, the contact structure of $J^{1}\bP$ defines a connection on the principal bundle  $J^{1}\bP \mapsto \mathcal{C}_{\bP}$ and this connection is universal in the sense that every principal connection on $\bP \mapsto \bX$ is induced by it via a section $\sigma: \bX \mapsto \mathcal{C}_{\bP}$. We will refer to it as the canonical connection $\phi$. 

Let $c^{k}_{ij}$ be the structural constants of $\mathfrak{g}$. We can write for the curvature of the canonical connection (see \eg \cite{CaMun01})
\beq
\mathcal{F} = \Sigma_{k} \, \mathfrak{g}_{k} \ten \, ( \, \Sigma_{\mu, \nu, \kappa} \, ( dx^{\mu} \wedge dA^{\stackrel{k}{\mu}} + \Sigma_{i, j} \, \frac{1}{2}c^{k}_{ij} \, A^{\stackrel{i}{\nu}}A^{\stackrel{j}{\kappa}} \, dx^{\nu} \wedge dx^{\kappa} ) ) \,.
\eeq
Its  horizontalization with respect to the contact structure on $J^{1}\mathcal{C}_{\bP}$ is then given by
\bEq \label{horF}
h(\mathcal{F}) = \Sigma_{k} \, \mathfrak{g}_{k} \ten \, ( \, \Sigma_{\mu, \nu} \, ( (A^{\stackrel{k}{\mu}}_{\,\, \nu} - A^{\stackrel{k}{\nu}}_{ \,\,\, \mu}) \, dx^{\mu} \wedge dx^{\nu}  + \Sigma_{i, j} \, \frac{1}{2}c^{k}_{ij} \, A^{\stackrel{i}{\mu}}A^{\stackrel{j}{\nu}} \, dx^{\mu} \wedge dx^{\nu} ) )\,
\eEq
with
\beq
(\pi_0^1)^*\mathcal{F} = \Theta_{\phi} + h(\mathcal{F}) \,.
\eeq
Here $\Theta_{\phi}$ is  of course the contact component of $(\pi_0^1)^*\mathcal{F}$; if $\theta^{\stackrel{k}{\mu}} = dA^{\stackrel{k}{\mu}} - \Sigma_{\nu} \, A^{\stackrel{k}{\mu}}_{\,\, \nu} \, dx^{\nu}$ are the local contact one forms on $J^{1}\mathcal{C}_{\bP}$, its local expression is given by
\bEq \label{theta}
\Theta_{\phi} = \Sigma_{k} \,\, \mathfrak{g}_{k} \ten \, \Sigma_{\mu} \,  dx^{\mu} \wedge \theta^{\stackrel{k}{\mu}}  \,.
\eEq
Now, if $(\pi_0^1)^* J^{1}\bP$ is the total space of the pullback of the principal bundle $J^{1}\bP \mapsto \mathcal{C}_{\bP}$ to $J^{1}\mathcal{C}_{\bP}$ via the jet bundle projection 
$\pi_0^1: J^{1}\mathcal{C}_{\bP} \mapsto \mathcal{C}_{\bP}$,
we can think either 
$\Theta_{\phi}, h(\mathcal{F}) \in \Lambda^2 (J^{1}\mathcal{C}_{\bP}) \ten V\bP / \bG$ or 
$ \Theta_{\phi}, h(\mathcal{F}) \in \Lambda^2 ( (\pi_0^1)^* J^{1}\bP) \ten \mathfrak{g}$, depending on the interpretation of $\mathcal{F}$.

We have the following expression for the curvature $\Omega_{\sigma}$ of the connection $\omega_{\sigma}$ with $\sigma(x) = (x, A^{\stackrel{k}{\mu}}(x) )$ 
\bEq \label{Omega}
\Omega_{\sigma} = (j^{1}\sigma)^{*}h(\mathcal{F})  = \sigma^{*} \mathcal{F} 
\eEq 
with coordinate expression
\beq
&\Omega_{\sigma} = \Sigma_{k}  \mathfrak{g}_{k} \ten 
\\ 
& \ten \,  [ \Sigma_{\mu, \nu}  (  (\frac{\der}{\der x_\nu}A^{\stackrel{k}{\mu}} (x) - \frac{\der}{\der x_\mu}A^{\stackrel{k}{\nu}}(x) ) dx^{\mu} \wedge dx^{\nu}  + \Sigma_{i, j} \frac{1}{2}c^{k}_{ij}  A^{\stackrel{i}{\mu}}(x)A^{\stackrel{j}{\nu}}(x)  dx^{\mu} \wedge dx^{\nu}  )  ]  \,.
\eeq

\subsection{Chern--Simons theories on the bundle of connections}

We will now outline the construction of classical Chern--Simons theories. Starting point is the work by Chern and Simons on secondary characteristic classes, \cite{ChSi71} and \cite{ChSi74}. Chern--Simons theories are gauge theories for connections. These gauge theories were introduced in \cite{DJT82} in three dimensions and in \cite{Ch89} and \cite{Ch90} in higher dimensions.
Interestingly, these higher dimensional theories have been fairly extensively studied in theoretical physics, see \eg \cite{Zan12}.

Let $\cP_{k}$ be the $k$-th $G$-invariant polynomial on $\mathfrak{g}$ the Lie algebra of $\bG$, let $\omega$ be a connection on the principal bundle $\bP$ over $\bX$ and $\Omega$ its curvature form
$\mathcal{P}_{k}(\Omega)$ is a closed form of degree $2k$ on $\bP$. Horizontal and invariant under the right action of $G$, it can also be considered a closed form on $\bX$. The cohomology class it defines is the $k$-th real valued Chern class of $P$. It is independent of the connection $\omega$ (this is the starting point of Chern-Weil-theory).

Chern and Simons found in \cite{ChSi74} a $2k-1$-form  $T\mathcal{P}_{k}(\omega)$ on $P$ such that $$\mathcal{P}_{k}(\Omega) = dT\mathcal{P}_{k}(\omega)$$ on $\bP$. One can write explicitly \cite{ChSi74}
\beq
T\mathcal{P}_{k}(\omega) = \Sigma_{i=0}^{k-1} \kappa_{i}\mathcal{P}_{k}(\omega \wedge [\omega, \omega]^{i} \wedge \Omega^{k-i-1})
\eeq
$\kappa_{i}$ is a rational numerical factor which is of no concern to us; note that we view $\cP_{k}$ here as a linear form on a $k$-th power of $\mathfrak{g}$, for more details see  again \cite{ChSi74}.

For every coordinate covering of $\bX$ over which $\bP$ can be trivialized and a corresponding family of local connection forms $\omega_U$, we get, thus, a family of local potentials $T\mathcal{P}_{k}(\omega_{U})$ for $\mathcal{P}_{k}(\Omega)$.

If $\mathcal{P}_{k}(\Omega)=0$ we have $0= dT\mathcal{P}_{k}(\omega)$ and, thus, $T\mathcal{P}_{k}(\omega)$ defines a cohomology class on $P$.
This cohomology class depends on the connection $\omega$. This dependence is the starting point of Chern--Simons-Field theories, the central idea is to use $T\mathcal{P}_{k}(\omega)$ as a variational principle for connections.\\
From now on we will consider a $U(n)$-principal bundle $\bP$ over a $(2p+1)$-dimensional manifold $\bX$ with $n \geq p+1$. We will also deal exclusively with the $p+1$-Chern polynomial $\cP_{p+1}$. At the end we will remark on some possible alternative choices.

We will sketch the construction of a local variational problem starting from the Chern--Simons form $T\mathcal{P}_{p+1}(\phi)$ for the canonical connection $\phi$ on the bundle  $J^{1}\bP \mapsto \mathcal{C}_{\bP}$. For a family of local connection forms $\phi_U$ on a covering of $\mathcal{C}_{\bP}$ with coordinate patches, we get, thus, a system of local potentials $T\mathcal{P}_{p+1}(\phi_{U})$ for $\mathcal{P}_{p+1}(\mathcal{F})$.
This system of local potentials projects now in the variational sequence on $J^{1}\mathcal{C}_{\bP}$ onto the system of local Lagrangians
\beq
\lambda^{CS}_{U} = \Sigma_{i=0}^{p} \kappa_{i}\mathcal{P}_{p+1}(\phi_{U} \wedge [\phi_{U}, \phi_{U}]^{i} \wedge (\Omega_{\Sigma}|_{U})^{p-i})
\eeq
Again we consider $\cP_{p+1}$ here to be linear on a power of $\mathfrak{g}$.
The dynamical form (or, {\em cum grano salis}, the Euler--Lagrange morphism) of this local variational problem is
\beq
\eta_{CS} = (\pi^{2}_{1})^{*}(\mathcal{P}_{p+1}((\Omega_{\Sigma})^{p}\wedge \mathcal{F}))
\eeq
This (once again we consider $\cP_{p+1}$ to be linear) is the image  (of the pullback to $J^{2}\mathcal{C}_{\bP}$) of
$
\mathcal{P}_{p+1}(\mathcal{F})
$ in the variational sequence.\\
In particular, $\mathcal{P}_{p+1}(\mathcal{F})$ and $\eta_{CS}$ define the same cohomology class in $H^{n+1} (J^{2}\mathcal{C}_{\bP})$. Since $H^{n+1} (J^{r}\mathcal{C}_{\bP}) \sim H^{n+1} (\mathcal{C}_{\bP}) \sim H^{n+1} (X) = 0$, the theory is globally variational, but a global Lagrangian is not easy to find and quite tricky to work with. It also seems to depend always on fixing a physical quantity, a background connection, {\em a priori}. See \cite{BFF03} for the $3$-dimensional case and \cite{GiMaSa03} for higher dimensions.

\section{The obstructions in the case of Chern--Simons theories for $U(p+1)$ and the $p+1$th Chern polynomial} \label{uni1}

Let $\bP \mapsto \bX$ be a  $U(p+1)$ principal bundle over the $(2p+1)$-dimensional manifold $\bX$.  Since $U(p+1)$ is a matrix group, it and its Lie algebra $\mathfrak{u}_{p+1}$ have standard representations. All matrix expressions, in particular determinants, are understood to be relative to these. 

Let $c_{k}$ be the  $k$th Chern polynomial on $\mathfrak{u}_{p+1}$ (see \eg \cite{KobNo} volume II, XII.$3$); for $A \in \mathfrak{u}_{p+1}$ it is defined by
\bEq \label{chernpol}
c_{k}(A) =  \frac{1}{(k)!}\left(\frac{i}{2\pi}\right)^{k} \cdot \Sigma_{i=1}^{p+1} \, M^k_i (A) 
\eEq
$\Sigma_{i=1}^{p+1} \, M^k_i (A) $ stands for the sum over all $k$-principal minors of $A$. If $\Omega$ is the curvature $2$-form of a connection $\omega$ on $\bP$, we have by Chern--Weil theory
\bEq \label{chernclass}
[c_{k}(\Omega)] = c^{\R}_{k}(\bP) \in H^{2k} (\bX, \R) \,
\eEq
where $c^{\R}_{k}(\bP)$ is the $k$th real Chern class of $\bP$ (for the integer Chern classes in algebraic topology see \eg \cite{Huse66} chapters $17$ and $20$ and \cite{MilSta} chapters $13$ to $16$). Note that $\frac{1}{(k)!}\left(\frac{i}{2\pi}\right)^{k}$ is a normalization factor such that $[c_{k}(A)]$ lies in the image of $H^{2k} (\bX, \Z)$ in $H^{2k} (\bX, \R)$ induced by the inclusion $\Z \subset \R$.
Thus, we have
\beq
c_{p+1}(\Theta_{\phi} + h(\mathcal{F})) =  \frac{1}{(p+1)!}\left(\frac{i}{2\pi}\right)^{p+1} \cdot  Det (\Theta_{\phi} + h(\mathcal{F})) 
\eeq
and 
\beq 
\eta_{CS} =c_{p+1}(\Theta_{\phi} \wedge h(\mathcal{F})^p) = 
\eeq
\bEq \label{minoreq}
\frac{1}{(p+1)!}\left(\frac{i}{2\pi}\right)^{p+1} \cdot \left(\Sigma_{pq} (-1)^{p+q} (\Theta_{\phi})_{pq} \wedge M_{pq}\left(h(\mathcal{F})\right) \right)
\eEq
$(\Theta_{\phi})_{pq}$ is the $pq$-entry of $\Theta_{\phi}$ and $M_{pq}\left(h(\mathcal{F})\right)$ is the $pq$-minor of $h(\mathcal{F})$, \ie the determinant of the 
matrix obtained by deleting the $p$th row and $q$th column of $h(\mathcal{F})$. Note, that the right hand expression is globally well defined on $(\pi_0^1)^* J^{1}\bP$, because there $h(\mathcal{F})$ and $\Theta_{\phi}$ are Lie algebra valued. 

\bPr \label{rank}
Let $\bP \mapsto \bX$ be a $U(p+1)$ principal bundle over a manifold $\bX$ of dimension $2p+1$. A principal connection $\omega$ on $\bP$ is a solution of the classical Chern--Simons gauge theory derived from the $(p+1)$th Chern polynomial if and only if the rank of its curvature form is everywhere $\leq p-1$.
\ePr
\bPf
 Solutions are sections $\sigma$ of $\pi_{\mathcal{C}_{\bP}}: \mathcal{C}_{\bP} \mapsto \bX$ such that $\eta_{CS} \, \circ \, j^1 \sigma = 0$. Keeping in mind that $\Theta_{\phi}$ does not vanish anywhere along any section , see (\ref{theta}) above, we see from (\ref{minoreq}) above that $M_{mn}\left(h(\mathcal{F})\right)|_{\sigma} = 0$. Since $\Omega_{\sigma} = (j^{1}\sigma)^{*}h(\mathcal{F})$, see (\ref{Omega}) above, this means that all $pq$-minors of $\Omega_{\sigma}$ vanish, \ie $\Omega_{\sigma}$ is everywhere of rank $\leq p-1$.
\ePf

With this characterization of the solutions at hand we can identify directly the $p$th real Chern class of $\bP$, $c^{\R}_{p}(\bP)$ as an obstructions to existence of global solutions.
\bCr
The vanishing of the $p$th real Chern class of $\bP$, $c^{\R}_{p}(\bP)$, is a necessary condition for the existence of global solutions in the situation of the above 
proposition. 

\eCr
\bPf
For a matrix $B \in \mathfrak{u}_{p+1}$, $c_p(B)$ is up to a normalization factor the sum over the $p$-principal minors of $B$ (see \eqref{chernpol} above) and
$c_p(B) = 0$ if and only if rank $B \leq p-1$.
By \eqref{chernclass} $c^{\R}_{p}(\bP) = [c_{p} (\Omega)]$ for any principal connection $\omega$ on $\bP$. But if $\omega_{\sigma}$ is a solution then rank $\Omega_{\sigma} \leq p-1$ by proposition \ref{rank} and we have $c_{p} (\Omega_{\sigma}) = 0$ and the $p$th real Chern class vanishes. Vice versa, if $c^{\R}_{p}(\bP)$ is non trivial, then we have $c_{p} (\Omega) \neq 0$ for all principal connection $\omega$ on $\bP$ and there cannot be a connection $\omega_{\sigma}$ with curvature $\Omega_{\sigma}$ of rank $\leq p-1$.
\ePf

We will now show that if $ 0 \neq c^{\R}_{p}(\bP) \in H^2 (\bX, \R)$ there is also a nontrivial cohomology class of type $[\Xi \rfloor \eta_{CS}]$ in the variational sequence, respectively a nontrivial cohomology class $[(j^1 \sigma)^* \Xi \rfloor \eta_{CS}] \in H^{2p+1}(\bX, \R)$.

Having already identified $c^{\R}_{p}(\bP)$ as obstruction to the existence of solutions, one naturally will wonder about  the relevance of this: a situation where the obstructions $\Xi \rfloor \eta_{CS}$ are easier accessible is hard to imagine. There are, however, three aspects which render this results quite interesting.

Firstly, its theoretical importance lies in the fact that it shows the power of the cohomological formulations of the calculus of variations: not only allows it to identify de Rham cohomology classes as obstruction to the existence of solutions, but it can do so by means intrinsic to the calculus of variations. In this sense this is a continuation and extension of the results of \cite{PaWi16}.

Secondly, as we mentioned already, the obstructions of type $\Xi \rfloor \eta_{CS}$ arise in the context of the Noether theorems in variational cohomology as obstructions to the existence of global conserved quantities  (see \eg \cite{FePaWi10}, \cite{PaWi16}, also \cite{And92}). Therefore, by making the connection between the two types of obstructions explicit, we identify the the $p$th real Chern class of $\bP$ as an obstruction to the globality of conserved quantities.

Thirdly, in the case of the coupling $\tilde{\lambda} +  \kappa \cdot \lambda_{CS}$ of the Chern--Simons Lagrangian  with another Lagrangian, we obtain the Euler--Lagrange form $\eta_{\tilde{\lambda}} + \kappa \cdot \eta_{CS}$. For the obstructions of the type $[\Xi \rfloor \left( \eta_{\tilde{\lambda}} + \kappa \cdot \eta_{CS} \right)]$ to vanish, this means then that $[\Xi \rfloor \eta_{\tilde{\lambda}}]$ is the precise counterpart which needs to annihilate $\kappa \cdot [\Xi \rfloor \eta_{CS}]$. This may be used to derive strong non existence theorems with direct applications in theoretical physics \cite{WYMCS}.

\bTh \label{obstruction2}
Let $\bP \mapsto \bX$ be an $U(p+1)$--principal bundle over the $2p+1$-dimensional closed manifold $\bX$; let $\eta_{CS}$ be the Euler--Lagrange form of the Chern--Simons gauge theory on $\bP$  derived from the $(p+1)$th Chern polynomial.

If the the $p$th real Chern class of $\bP$, $c^{\R}_{p}(\bP)$ does not vanish then there exists a vertical vector field $\Xi$ on the respective bundle of connections $\mathcal{C}_{\bP} \mapsto \bX$ such that the corresponding obstruction of the form $[\Xi \rfloor \eta_{CS}]$ does not vanish either. 
Conversely, if $[\Xi \rfloor \eta_{CS}]$ is nontrivial, so is $c^{\R}_{p}(\bP)$.
Furthermore, we have 
\beq
\Xi \rfloor \eta_{CS} = h\left(\frac{-1}{2\pi \cdot (p+1)} \cdot \pi^*(\beta) \, \wedge c_p\left(\mathcal{F}\right)\right)
\eeq 
and
\beq
(j^1 \sigma)^* \Xi \rfloor \eta_{CS} = \frac{-1}{2\pi \cdot (p+1)} \cdot \beta \wedge c_p\left(\Omega_{\sigma}\right)
 \eeq
where $\beta$ is a differential $1$-form on $\bX$.
\eTh
\bPf
For the sake of notational simplification we set $\lambda_k:= \frac{1}{(k)!}\left(\frac{i}{2\pi}\right)^{k}$.
From equation \eqref{theta} we get locally 
\beq 
\Theta_{\phi} = \Sigma_{k} \,\, \mathfrak{g}_{k} \ten \, \Sigma_{\mu} \,  dx^{\mu} \wedge \left(dA^{\stackrel{k}{\mu}} - \Sigma_{\nu} \, A^{\stackrel{k}{\mu}}_{\,\, \nu} \, dx^{\nu} \right)
\eeq
and, thus, with $\Xi = \Sigma_{k, \mu} \, \xi^{\stackrel{k}{\mu}}  \frac{\der}{\der A^{\stackrel{k}{\mu}}}$ we have
\beq
\Xi \rfloor \eta_{CS} =  \Xi \rfloor \left( \lambda_{p+1} \cdot \Sigma_{pq} (-1)^{p+q} (\Theta_{\phi})_{pq} \wedge M_{pq}\left(h(\mathcal{F})\right)\right) =
\eeq
\beq
\lambda_{p+1} \cdot \Sigma_{pq} (-1)^{p+q} (\Sigma_{k} \,\, \mathfrak{g}_{k} \ten \, \Sigma_{\mu} \, \xi^{\stackrel{k}{\mu}}  dx^{\mu})_{pq} \wedge M_{pq}\left(h(\mathcal{F})\right)
\eeq
For a section $\sigma: \bX \mapsto \mathcal{C}_{\bP}$, the corresponding connection $\omega_{\sigma}$ and its curvature $\Omega_{\sigma}$, we have 
\beq
(j^1 \sigma)^* \Xi \rfloor \eta_{CS} = \lambda_{p+1} \cdot \Sigma_{pq} (-1)^{p+q} (\Sigma_{k} \,\, \mathfrak{g}_{k} \ten \, \Sigma_{\mu} \, \xi^{\stackrel{k}{\mu}}\left(\sigma(x)\right) dx^{\mu})_{pq} \wedge M_{pq}\left(\Omega_{\sigma}\right)
\eeq
As the next step, we need to construct a particular vertical vector field $\Xi$.
To this end note that, since $\pi_{\mathcal{C}_{\bP}}: \mathcal{C}_{\bP} \mapsto \bX$, is an affine bundle modeled on the vector bundle 
\beq
T^* \bX \ten  V\bP/U(p+1) \mapsto \bX
\eeq
we have
\beq
V\mathcal{C}_{\bP} \, \sim \, \mathcal{C}_{\bP} \, \times_{\bX} \, T^* \bX \ten V\bP /U(p+1)
\eeq
and, in particular, we can identify
\beq
\frac{\der}{\der A^{\stackrel{k}{\mu}}} \, = \, \mathfrak{g}_{k} \ten dx^{\mu}
\eeq
Since $\left(i \cdot  \bf{1}_{p+1}  \right) \in \mathfrak{u}_{(p+1)} $ is $\,U(p)$--invariant, $\left(i \cdot  \bf{1}_{p+1}  \right) \ten \beta$ defines a unique vertical vector field $\Xi$ on $\mathcal{C}_{\bP}$ for any  differential $1$-form $\beta$ on $\bX$. If we identify $\left(i \cdot  \bf{1}_{p+1}  \right) \in \mathfrak{u}_{(p+1)} $ with $\mathfrak{g}_{0}$ and with the local expression $\beta = \Sigma_{\mu} \xi^{mu}\cdot dx^{\mu}$, we have locally $\Xi = \Sigma_{\mu} \, \xi^{\mu}  \frac{\der}{\der A^{\stackrel{0}{\mu}}}$. Therefore, we have
\beq
\Xi \rfloor \eta_{CS} = \lambda_{p+1} \cdot \Sigma_{p} \, i \cdot \beta \wedge M_{pp}\left(h(\mathcal{F})\right) = 
\eeq
\beq
\frac{-1}{2\pi \cdot (p+1)} \cdot (\pi_0^1)^*(\beta) \, \wedge \left(\lambda_{p} \cdot \Sigma_{i=1}^{p+1} \, M^p_i \left(h(\mathcal{F})\right)\right) =
\eeq
\beq
\frac{-1}{2\pi \cdot (p+1)} \cdot (\pi_0^1)^*(\beta) \, \wedge c_p\left(h(\mathcal{F})\right)
\eeq
on $(\pi_0^1)^* J^{1}\bP$ by equation (\ref{minoreq}) and, thus, on $\bX$
\beq
(j^1 \sigma)^* \Xi \rfloor \eta_{CS} = \frac{-1}{2\pi \cdot (p+1)} \cdot \beta \wedge c_p\left(\Omega_{\sigma}\right)
 \eeq
If now $0 \neq c^{\R}_{p}(\bP) \in H^{2p}(\bX, \R)$ Poincar\'e duality implies that $\beta$ can be chosen such that 
\beq
0 \neq [(j^1 \sigma)^* \Xi \rfloor \eta_{CS}] = \frac{-1}{2\pi \cdot (p+1)} \cdot [\beta] \cup c^{\R}_{p}(\bP) \in H^{2p+1}(\bX, \R) \,
\eeq
 since $\bX$ is closed. Thus, it only remains to be shown that $\left(\left(i \cdot  \bf{1}_{p+1}  \right) \ten \beta\right) \rfloor \eta_{CS}$ defines a cohomology class in the variational sequence. 
To see this, note first that we have
\bEq \label{Conv}
\frac{-1}{2\pi \cdot (p+1)} \cdot \beta \wedge c_p\left(\Omega_{\sigma}\right) = \sigma^* \left( \frac{-1}{2\pi \cdot (p+1)} \cdot \pi^*(\beta) \, \wedge c_p\left(\mathcal{F}\right)\right)
\eEq
Now, 
$\frac{-1}{2\pi \cdot (p+1)} \cdot (\pi_0^1)^*(\beta) \, \wedge c_p\left(\mathcal{F}\right)$  represents a nontrivial cohomology class whenever $(j^1 \sigma)^* \Xi \rfloor \eta_{CS}$ does,
since $[c_p\left(\mathcal{F}\right)] = c^{\R}_{p}(J^1\bP)$. For $\Xi = \left(i \cdot  \bf{1}_{p+1}  \right) \ten \beta$ we have then
\beq
\Xi \rfloor \eta_{CS} = h\left(\frac{-1}{2\pi \cdot (p+1)} \cdot \pi^*(\beta) \, \wedge c_p\left(\mathcal{F}\right)\right)
\eeq 
But since the horizontalization $h$ is the projection onto the variational sequence for $2p+1$-forms, we have shown that there is a cohomological non trivial $\Xi \rfloor \eta_{CS}$, whenever $c^{\R}_{p}(\bP)$ is cohomological non trivial. 

For the converse statement, note that $[\Xi \rfloor \eta_{CS}]$ is the image of $[\frac{-1}{2\pi \cdot (p+1)} \cdot \pi^*(\beta) \, \wedge c_p\left(\mathcal{F}\right)] \in H^{2p+1}(\mathcal{C}_{\bP}, \R)$ in the cohomology of the variational sequence; \ie the former is non trivial if and only if the latter is. For the latter we have
\beq
[\frac{-1}{2\pi \cdot (p+1)} \cdot \pi^*(\beta) \, \wedge c_p\left(\mathcal{F}\right)] = \frac{-1}{2\pi \cdot (p+1)} \cdot \pi^*[\beta] \, \cup [c_p\left(\mathcal{F}\right)]
\eeq
Since $\sigma^*$ and $\pi^*$ are inverse isomorphisms in cohomology, we get from equation (\ref{Conv})
\beq
\frac{-1}{2\pi \cdot (p+1)} \cdot \pi^*[\beta] \, \cup [c_p\left(\mathcal{F}\right)] = \frac{-1}{2\pi \cdot (p+1)} \cdot \pi^*[\beta] \, \cup \pi^*[c_p\left(\Omega_{\sigma}\right)] = 
\eeq
\beq
\frac{-1}{2\pi \cdot (p+1)} \cdot \pi^*[\beta] \, \cup \pi^* c^{\R}_{p}(\bP)
\eeq
Therefore, if $[\Xi \rfloor \eta_{CS}]$ is nontrivial, $c^{\R}_{p}(\bP)$ also is (and $[\beta]$ is up to a constant multiple its Poincar\'e dual).
\ePf

\section*{Acknowledgements}
Research partially supported by Department of Mathematics - University of Torino research project PALM$\_$RILO$\_20\_ 01$. 
The first author (MP) also would like to acknowledge the contribution of the COST Action CA17139. The second author (EW) acknowledges membership in the Lepage Research Institute, Pre\v sov, Slovakia.


\end{document}